%% file: aipsamp.tex
\documentclass[%
 aip,
 amsmath,amssymb,
 reprint,%
]{revtex4-1}

\usepackage{graphicx}% Include figure files
\usepackage{dcolumn}% Align table columns on decimal point
\usepackage{bm}% bold math
\usepackage[utf8]{inputenc}
\usepackage[T1]{fontenc}
\usepackage{mathptmx}
\usepackage{etoolbox}
\usepackage{color}
\usepackage{hyperref}
\usepackage{tabularx}
\usepackage{makecell}
\usepackage[figuresleft]{rotating}

\include{vpack}

\makeatletter
\def\@email#1#2{%
 \endgroup
 \patchcmd{\titleblock@produce}
  {\frontmatter@RRAPformat}
  {\frontmatter@RRAPformat{\produce@RRAP{*#1\href{mailto:#2}{#2}}}\frontmatter@RRAPformat}
  {}{}
}%
\makeatother
\begin{document}

\preprint{AIP/123-QED}

%\title{Data-driven subgrid closures for Burgers' turbulence using differentiable physics}
\title{Differentiable physics-enabled closure modeling for Burgers' turbulence}
%\title{End-to-end learning of closure models with differentiable physics for Burgers' turbulence}

\author{Varun Shankar}
\affiliation{Carnegie Mellon University}
%\email{varunshankar@cmu.edu} 

\author{Vedant Puri}
\affiliation{Carnegie Mellon University}
%\email{vedantpuri@cmu.edu} 

\author{Ramesh Balakrishnan}
\affiliation{Argonne National Laboratory}
%\email{bramesh@anl.gov} 

\author{Romit Maulik}
\affiliation{Argonne National Laboratory}
\email{rmaulik@anl.gov} 

\author{Venkatasubramanian Viswanathan}
\affiliation{Carnegie Mellon University}
\email{venkvis@cmu.edu}

\date{\today}

\begin{abstract}
    Data-driven turbulence modeling is experiencing a surge in interest following algorithmic and hardware developments in the data sciences. We discuss an approach using the differentiable physics paradigm that combines known physics with machine learning to develop closure models for Burgers' turbulence. We consider the 1D Burgers system as a prototypical test problem for modeling the unresolved terms in advection-dominated turbulence problems. We train a series of models that incorporate varying degrees of physical assumptions on an \textit{a posteriori} loss function to test the efficacy of models across a range of system parameters, including viscosity, time, and grid resolution. We find that constraining models with inductive biases in the form of partial differential equations that contain known physics or existing closure approaches produces highly data-efficient, accurate, and generalizable models, outperforming state-of-the-art baselines. Addition of structure in the form of physics information also brings a level of interpretability to the models, potentially offering a stepping stone to the future of closure modeling. 
\end{abstract}


\maketitle

\section{Introduction}
\label{sec:intro}

% need for closure modeling
Simulation of fluid flow for practical applications is characterised by high Reynolds number turbulent flows over complex geometries\cite{homiff,karniadakis1999simulating}. The temporal and spatial scales of motion in such flows span several orders of magnitude\cite{moin_DNS_research}. The cost of Direct Numerical Simulations (DNS), which attempt to capture all energy containing lengthscales in the flow, grows superlinearly with the Reynolds number of the flow\cite{kolmogorov, dns_primer, orszag_1970}, limiting the scope of fully resolving simulations to canonical geometries at medium Reynolds numbers\cite{moin_DNS_research}.

% closure modeling
Turbulence closure models have been developed for cases where resolving and evolving the entirety of velocity and pressure spectra is not possible. Techniques such as Large Eddy Simulations (LES) and Reynolds Averaged Navier-Stokes (RANS) allow for cheaper calculations of high Reynolds number flows by resolving only a portion of the spectrum and `modeling' the rest \cite{pope}. Put simply, the goal of turbulence modeling is to find a closed system that can describe the time-evolution of observables, such as the resolved velocity spectrum, via a `closure' map that allows one to untangle the dependence of said observables on unresolved variables\cite{sagaut2006large,adams2007mathematics,ml_stat_closure}.

% intro to ML
The problem of turbulence closure modeling has been extensively studied for over a century, dating back to Boussinesq's eddy viscosity hypothesis in 1877\cite{Layton2014THE1B}. More recently, machine learning (ML) based methods have utilized for modeling fluid flow problems\cite{Brunton2020,vinuesa2021potential,turb_modeling_data,kochkov2021machine,watt2021correcting,Vlachas2018,Thuerey_2020,portwood2021interpreting,Wang2020}. The success of machine learning is built upon the ability of deep neural networks to approximate functions in high-dimensional spaces with a number of parameters that does not scale exponentially with the dimension of the problem space\cite{weinan}. As the underlying problem in closure modeling is of finding the functional mapping between high-dimensional vector spaces,
% ie from the space of resolved velocity and pressure on a computational grid, $\mathbb{R}^{4N}$ where the mesh has $N$ points, to itself,
deep neural networks can be utilized in a supervised learning framework.

% ML + closure lacking
However, the performance of machine learning models to turbulence modeling has largely been unsatisfactory. A probable reason is that while neural networks are touted to break the curse of dimensionality, the cost of attaining an accurate model may be larger. The cost of generating high-fidelity data for the model does scale with dimensionality\cite{pope}, and even if data is readily available, machine learning models need to be trained on large number of samples to accurately approximate even simple functions. In other words, an optimal closure model may be present in the space of functions that a neural network architecture can express, but there is no clear path towards realizing it.
%model data inconsistnency - spurious wave modes, instabilities

Some success has been seen, however, by augmenting neural network models with physical priors. For example, Ling et al.\cite{ling_kurzawski_templeton_2016,milani2021} improved upon conventional machine learning models by embedding Galilean invariance in the prediction of Reynolds stress tensor for Reynolds Averaged Navier-Stokes simulations. Shankar et al.~\cite{shankar2020learning} enhanced the output of a convolutional neural network model on homogeneous isotropic turbulence by imposing the divergence-free constraint on the model output. Both approaches can be understood as attempts to narrow down the space of functions that a neural network architecture can represent.

% diff phys
Rackauckas et al.\cite{universal_diffeq} further develops inductive biases in data-driven learning by directly embedding trainable parameters in a model's governing differential equations. Training such parameters against a loss function that depends on the solution to the governing system requires passing gradient information through a differential equation solve. Therefore a solver that is compatible with automatic-differentiation toolchains in needed. Directly backpropagating gradients is computationally expensive for large solves, so the adjoint optimization method\cite{Chen,farrell2013automated,errico1997adjoint,sigmund2013topology} is utilized. CFD codes such as SU2 have utilized the adjoint method for sensitivity analysis and shape optimization\cite{su2}.

\begin{figure*}
  \centering
  \includegraphics[width=0.8\linewidth]{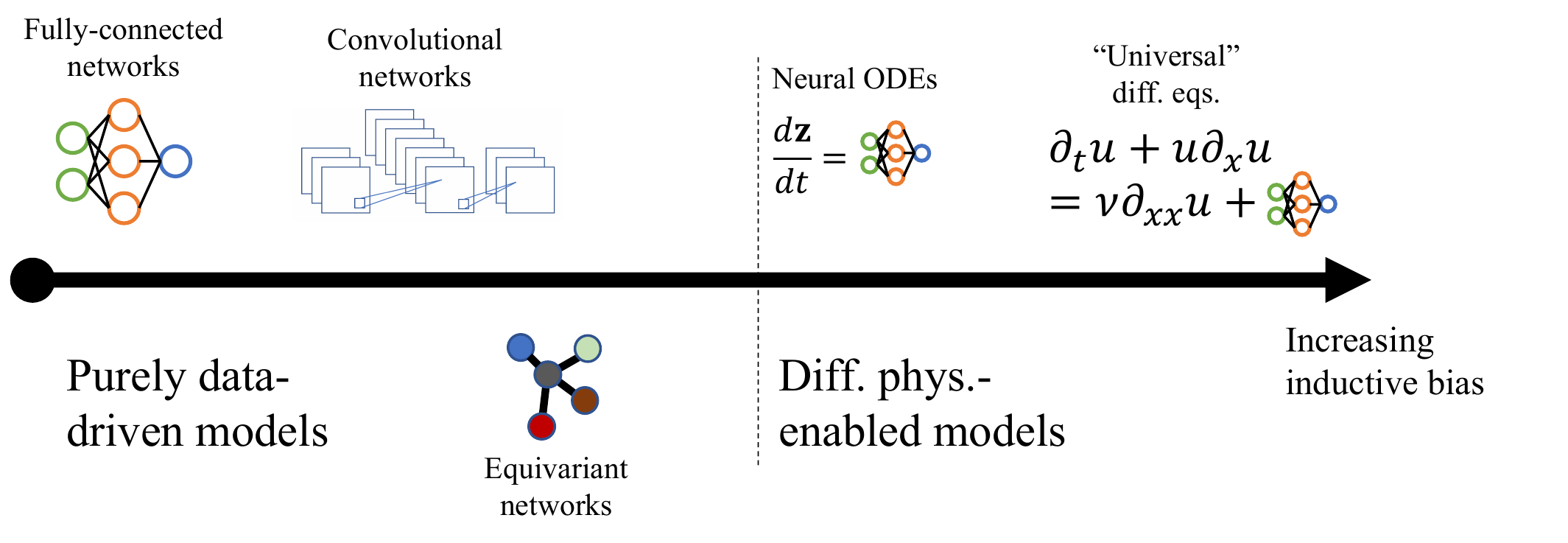}
  \caption{Data-driven models can incorporate different inductive biases, from locality in convolutional networks, symmetries in equivariant networks, to differential equation forms in universal differential equations. Models on the right are constrained by known physics, such as a dynamical system or advective and dissipative terms, leading to interpretable models where the learning problem has been restricted to unknown or residual physics.}
  \label{fig:dp}
\end{figure*}

This idea is formalized into the notion of Differentiable Physics\cite{diffphys}, defined as the set of scientific avenues emerging from the marriage of state-of-the-art partial differential equation (PDE) solvers with differentiable programming. The promise of differentiable physics is that machine learning models can be developed to learn specific unknown or residual physics, leading to more interpretable models than classical ML approaches. We envision the landscape of data-driven models on an axis of increasing inductive biases in \autoref{fig:dp}.

% fourier neural operators
%\textcolor{orange}{Vedant: few words on generalizing over resolutions?}
%Another roadblock in the development of generalized machine learning closures is that neural networks trained on a set computational grid do not generalize to other grids. This is because such networks learn function mappings between two set Euclidean spaces, and the implicit dependence on the computational grid they were trained on. Resolution invariant architectures, such as the Fourier Neural Operator\cite{fno}, have been recently developed that parameterize the computational kernel over some latent space that can be transformed to family of computational grids. \textcolor{blue}{Varun: might be something to trim if needed}

% this work
In this work, we comprehensively test the ability of ML techniques to develop turbulence closure models that are resolution independent, interpretable in the language of flow physics, and generalizeable across a range of Reynolds numbers. We consider the 1D Burgers system as the ideal test problem for modeling the energy spectra observed in advection-dominated turbulence problems, and narrow our focus to the task of subgrid-stress modeling. We test a range of differentiable physics models, that vary from a black-box approach to training coefficients in state-of-the-art turbulence models. Experiments are conducted over a range of Reynolds numbers and grid resolutions to assess the generalizability of the model, and over multiple resolution-independent architectures. We find that physics-based inductive biases are critical to the development of data-efficient and accurate ML closures, and that these ML closures can outperform state-of-the-art closure baselines in a wide range of flows.

\section{Background}
\label{sec:background}

\subsection{Burgers Turbulence}

% energy cascade
Turbulent flows are characterised by a competition between viscous forces, which damp out velocity fluctuations by converting kinetic energy into heat, and inertial forces, such as gravitation or pressure gradients. These tend to generate and preserve velocity and pressure fluctuations\cite{wall-bounded-turb}. In low Reynolds number flows, viscous forces dominate and the flow is near perfectly damped, meaning that any arbitrary fluctuation in the velocity field would be smoothed out in no time\cite{damped}. The velocity fields adjusts almost instantaneously to any changes in the pressure gradient that drives the flow. At high Reynolds numbers, however, viscous forces may not be strong enough to dampen out velocity fluctuations, and even tiny disturbances may cause the entire flow to destabilise into turbulence\cite{instabilitiy}. Energy dissipation primarily happens at the smaller scales as the rate of dissipation scales with the wavenumber squared.

We restrict the scope of this work to the task of subgrid stress modeling in advection-dominated turbulence problems.
%As such, we seek to insulate ourselves from all other sources of complexity and nuance in typical CFD simulations. 
%Training a machine learning model on complicated system makes it hard to generalize as a machine learning model implicitly learns a dependence on the specificity of the dataset. As such we seek to simplify our target problem as much as possible without 
The unforced, viscous Burgers problem, \autoref{eqn:burg}, an advection-diffusion type problem\cite{burg} that exhibits an energy cascade similar to the Navier-Stokes system, is the perfect test bench. The Burgers system is governed by
\eqn{\label{eqn:burg}
    \partial_t u + u\partial_x u = \nu \partial_{xx}u,
}
where $u(x,t)$, the solution to the Burgers system, represents a time-varying velocity field and $\nu$ is the kinematic viscosity. The 1D viscous Burgers system is a canonical PDE that is widely used as a simplified case study for subgrid model development and analysis \cite{falkovich2006lessons,burg,love1980subgrid,labryer2015framework}. The nonlinear advective term coupled with viscous dissipation leads to similar multiscale phenomena observed in more complex flow dynamics. \autoref{eqn:burg} is deterministic which allows us to directly compare trajectories rather than statistical averages, speeding up model training. We insulate the problem from arbitrary effects of boundary conditions by considering \autoref{eqn:burg} on the entire real line, $\R$, which is approximated by periodic boundary conditions on an interval. The Burgers equation can also be written in its conservative form
\begin{equation}
    \label{eqn:burg-conservative}
    \partial_{t} u + \frac{1}{2}\partial_{x} u^2 = \nu\partial_{xx}u , 
\end{equation}
which is typically more amenable to numerical calculation.

\subsection{Large Eddy Simulations}

% LES
As high Reynolds number flows have a wide energy spectrum, DNS of such flows is often computationally intractable, and a turbulence model is needed. Large Eddy Simulations (LES)  resolves the flow variables starting from the intertial range up to a cut-off length, $\Delta$, that depends upon the computation grid\cite{lesieur1996new,boris1992new}. Critically, an LES calculation ignores the smallest lengthscales in the flow, which are most computationally expensive to resolve. Filtering is accomplished by 
\eqn{
    \label{eqn:filter}
    u(\vect{x}, t) &= \bar{u}(\vect{x}, t) + u'(\vect{x}, t)\\
    \bar{u}(\vect{x}, t) &= G_\Delta \star u(\vect{x},t),
}
where $G_\Delta$ is a low-pass filtering operator. Quantities of interest are interpolated onto a computational grid that can resolve flow features only up to lengthscale $\Delta$. The grid has much lower resolution than what is needed to fully resolve the flow, and evolving flow variables on this grid lead to saving computational resources. 

Filtering the Burgers equation in this manner leads to
\eqn{\label{eqn:burg-LES}
    \partial_t \bar{u} + \frac{1}{2}\partial_x \bar{u}^2 = \nu \partial_{xx} \bar{u} - \frac{1}{2}\partial_x\bar{u'^2},
}
the LES equations for the Burgers system. Evolving $\bar{u}$ per \autoref{eqn:burg-LES} is not possible as one does not have access to the subgrid stress term, ${\eta}\defeq\frac{1}{2}\bar{{u}'{u}'}$, and ignoring it leads to an incorrect flow field and numerical instabilities. For the Navier-Stokes equations, ignoring subgrid stresses leads to numerical instabilities spurious high-frequency modes\cite{fischer-mullen-filter}. The closure problem of LES is of modeling the effect of ${\eta}$ on the dynamics of $\bar{u}$. This is called subgrid stress modeling. As one does not have access to the unresolved quantities, subgrid stress models are formulated in terms of the resolved quantities.

Kolmogorov hypothesized that the finer scales are largely similar\cite{kolmogorov}, and are primarily responsible for energy dissipation. One approach to closure modeling is to take out small amounts of energy from the high-frequency modes in $\bar{{u}}$ by applying a mild low-pass filter at every timestep. Fisher et al.\cite{fischer-mullen-filter} has applied this methodology to the Navier-Stokes equations in the Computational Fluid Dynamics (CFD) code \textsc{NEK5000}\cite{nek5000}, and Layton et al.\cite{layton2010temporally} has prove existence of unique strong solutions to the resulting continuum model for the same.

Another approach is to directly model the commutative error term, ${\eta}$ in terms of $\bar{{u}}$ with a map $\eta(\bar{u})$, and evolve the flow variables as per \autoref{eqn:burg-LES}. Hypothesizing a functional form for additive closures of this sort is highly problem dependent. A common hypothesis is the eddy viscosity family of models\cite{frisch1995turbulence,SCHMITT2007617} that introduce an ``effective'' or ``eddy'' viscosity, $\nu_t$, to the momentum equation for enhancing dissipation. This leads to a functional form of ${\eta} = \partial_x\nu_t\partial_x\bar{u}$, where $\nu_t$ could be a fixed scalar based on known flow physics and computational grid, or be defined by a transport equation. In \autoref{sec:experiment} we consider several such models and apply them to the Burgers problem.
\section{Method}
\label{sec:method}

% We aim to quantify our approach to learning subgrid models by examining the one-dimensional viscous Burgers equation, given by the governing equation:
% \begin{equation}
%     \partial_{t} u + u\partial_{x} u = \nu\partial_{xx}u , 
% \end{equation}
% where $u(x,t)$, the solution to the Burgers system, represents a time-varying velocity field and $\nu$ is the kinematic viscosity. The 1D viscous Burgers system is a canonical partial differential equation (PDE) that is widely used as a simplified case study for subgrid model development and analysis \cite{burg}. The nonlinear advective term coupled with viscous dissipation leads to similar multiscale phenomena observed in more complex flow dynamics. The Burgers system can also be written in its conservative form:
% \begin{equation}
%     \partial_{t} u + \frac{1}{2}\partial_{x} u^2 = \nu\partial_{xx}u , 
% \end{equation}
% which is typically more amenable to numerical computation.

\subsection{Numerical methods}
We compute solutions to the Burgers equation via a pseudospectral method \cite{pseudospectral}. The periodic 1D domain is discretized with a regular grid of $N$ grid points. The periodic nature of the solutions allow for expansion in a Fourier basis, as per
\begin{equation}
    u(x,t)=\sum_{k=-N/2}^{N/2}\hat{u}_k(t)e^{ikx} .
\end{equation}

The linear term of the equation can be computed pointwise in Fourier space
\begin{equation}
\partial_{xx}\hat{u}(k,t) = -k^2\hat{u}(k,t) ,
\end{equation} 

however the nonlinear term introduces interactions between all the wavemodes and thus requires more care for efficient computation. % do we want more details
Therefore, it is more suitable to compute the quadratic term $u^2$ pointwise in real space at the $N$ collocation points, then compute the spatial derivative in Fourier space. %aliasing?

The spatial discretization allows for a method of lines approach to time evolution. The discrete Fourier modes from the spectral basis expansion produces a system of $N$ ordinary differential equations (ODEs) that can be solved with standard ODE integration schemes. We choose the Tsitouras adaptive 5th order scheme \cite{tsit5}.

\subsection{Problem statement}
Accurate simulation of the governing Burgers equation imposes certain restrictions on the temporal and spatial discretization. As we use an adaptive time-stepping scheme with fixed error tolerance, we focus on errors arising from the spatial discretization. The spatial grid should be sufficiently dense to resolve the smallest dissaptive length scales in the flow, typically much smaller than the characteristic length scales of the flow. 
%Our ground truth data is generated on such a grid, with $N=4096$ for a periodic domain of size $L=2\pi$. put in exp.

We wish to approximate solutions to the Burgers equation on a considerably coarser grid of $M$ points, by exploring a family of data-driven models enabled by algorithmic differentiation of the temporal and spatial discretization schemes. Each of the models is described by a PDE or set of PDEs, where certain terms in the equations are computed by deep neural networks. Thus, we enable significant interpretability of the models as network outputs can be readily identified as components of traditional closure modeling approaches.

\subsection{Models}
In this section, we provide a description of the various models that we use in our experiments. First, we make clear our notational conventions.

$u(x,t)$ is the spatiotemporal velocity field we are modeling. $\bar u(x,t)$ is the filtered velocity field on the LES grid. We drop the parentheses for brevity. $\eta$ represents an unknown closure variable we are modeling. The exact description of this variable differs across models. In some models, an additional transport equation is introduced to govern the dynamics of the closure variable. The  $ \overset{\bullet}{} $ and the $\nabla$ symbols correspond to the time derivative and spatial derivative of the field respectively. A subscript 0 represents the initial condition of the field.

Neural networks are indicated with subscript $\theta$. These functions are parameterized by a large number of weights with functional form described in the subsequent section. Network inputs and outputs are written as:
\begin{equation}
    c,d=f_\theta(a,b;e) , 
\end{equation}
where $a$ and $b$ are explicit inputs to the network, and $c$ and $d$ are outputs of the network with implicit dependence on variable $e$. For example, $\eta=f_\theta(u;x)$ represents a network that takes the spatially varying velocity field as input and produces a spatially varying $\eta$ field as output.

The ground truth equation, or \textit{none} model, is given by:
\begin{equation}
    \dot{\bar{u}} = \nu\nabla^2\bar u - \bar u\nabla \bar u
\end{equation}
Since there is no closure term or trainable networks, we use this as a baseline comparison for our learned models.

We also compare our models to a couple baseline closure models \cite{Maulik2018}, the constant Smagorinsky model (further denoted as \textit{smag-const}), and the dynamic Smagorinsky model (further denoted as \textit{smag-dyn}).
The constant Smagorinsky model\cite{smagorinsky1963general} is given by: %cite
\begin{equation}
    \dot{\bar{u}} = \nu\nabla^2\bar u - \bar u\nabla \bar u + \nabla(\nu_t\nabla \bar u)
\end{equation}
\begin{equation}
    \nu_t = (C_s\delta x)^2|\nabla \bar u| ,
\end{equation}
where $\nu_t$ is the eddy viscosity, $C_s$ is the Smagorinsky constant, and $\delta x$ is the grid spacing. While there are no trainable networks in this model, $C_s$ can be optimized via hand tuning or gradient descent.

The dynamic Smagorinsky model \cite{germano1991dynamic,lilly1992proposed} is similar to the constant model, however the term $(C_s\delta x)^2$ is determined from the state of the system. Studies have shown that $C_s$ is not constant and may be variable with different flow characteristics \cite{smagorinsky1993some}. The model uses a second test filter, denoted by $\sim$ with filter scale $\tilde{\delta} x$. We fix the filter ratio $\kappa=\tilde{\delta} x/\delta x =2$. Specifically,
\begin{equation}
    (C_s\delta x)^2 = \frac{\langle HM \rangle}{\langle M^2 \rangle} ,
\end{equation}
\begin{equation}
    H = \nabla(\widetilde{\bar{u}}^2/2) - \nabla(\widetilde{\bar{u}^2}/2) ,
\end{equation}
\begin{equation}
    M = \kappa^2 \nabla(|\nabla \widetilde{\bar{u}}|\nabla \widetilde{\bar{u}}) - \nabla(\widetilde{|\nabla \bar{u}|\nabla \bar{u}}) ,
\end{equation}
where $\langle \rangle$ denotes averaging over the spatial domain.

%\textcolor{red}{Romit: General comment - assuming a submission to a physics-venue it would be best to add one paragraph that explains the neural ODE here (for example how its trained, how the training data is collected).} \textcolor{blue}{Varun: @vedant can you include a high-level description of NODE principles in the intro/background? We go into the methodology in the optimization section}\\

The rest of the models in this section contain trainable networks in functional forms that rely on varying degrees of assumptions typically used in closure analysis. At one extreme, we have model \textit{resnet}, which uses the classical ResNet architecture \cite{resnet} to predict the coarse-grained flow field at the next time step:
\begin{equation}
    (\bar{u}_{t+\Delta t}, \eta_{t+\Delta t}) = (\bar{u}_{t}, \eta_{t}) + f_\theta(\bar u_t,\eta_t,\nu;x)\Delta t,
\end{equation}
\begin{equation}
    \eta_0 = g_{\theta}(\bar u_0; x),
\end{equation}
where $\Delta t$ is a coarse time step. This model makes no assumptions about the physical description of the system, including the fact that it may be described as a PDE. The closure variable $\eta$ is used only to lift the model into a higher-dimensional space governed by the neural network. The network may choose to learn $\eta_t=0$, or it may leverage $\eta$ in the computation of $\bar{u}_t$. Projecting or transforming an input into a higher-dimensional subspace is a common practice in ML, where a large latent space of hidden channels eases learning. This is a core component of augmented neural ODEs \cite{anode}, which have shown to generalize better than standard neural ODEs.

\textit{anode}, an augmented neural ODE, is a small modification to \textit{resnet} that assumes an ODE inductive bias. The model is given by:
\begin{equation}
    \dot{\bar{u}}, \dot \eta = f_\theta(\bar u,\eta,\nu;x)
\end{equation}
\begin{equation}
    \eta_0 = g_{\theta}(\bar u_0; x),
\end{equation}
where $f_\theta$ explicitly describes the RHS of an ODE governing the discretized flow field. The dynamics of $\bar u$ and $\eta$ are determined purely by the network $f_\theta$ with no additional terms. This involves only a small change in the code of \textit{resnet}, changing the solver from explicit Euler to an adaptive scheme.

Filtering the governing equations as in \autoref{eqn:burg-LES} introduces an additional closure term to the equations. Model \textit{direct} models the entirety of this term directly as an additive correction:
\begin{equation}
    \dot{\bar{u}} = \nu\nabla^2\bar u - \bar u\nabla \bar u + \eta 
\end{equation}
\begin{equation}
    \eta = f_\theta(\nabla \bar u, \nu; x).
\end{equation}

Model \textit{transport-I} leverages the fact that the closure term can be written as the divergence of an unknown subgrid stress. Here, $\eta$ represents this stress analogue in 1D, which is additionally transported with another dynamical equation:
\begin{equation}
    \dot{\bar{u}} = \nu\nabla^2\bar u - \bar u\nabla \bar u + \nabla\eta
\end{equation}
\begin{equation}
    \dot \eta = \alpha \cdot \nu\nabla^2\eta - \beta \cdot \bar u\nabla\eta
\end{equation}
\begin{equation}
    \alpha, \beta = f_\theta(\nabla \bar u, \nabla \eta, \nu; x)
\end{equation}
\begin{equation}
    \eta_0 = g_{\theta}(\bar u_0, \nu; x).
\end{equation}
The coefficients of the dissipative and convective terms in the $\eta$ transport equation are given by neural networks.

Model \textit{transport-II} uses the Boussinesq hypothesis that the subgrid stress has a linear relationship with the strain in the flow. $\eta$ represents an eddy viscosity analogue:
\begin{equation}
    \dot{\bar{u}} = \nu\nabla^2\bar u - \bar u\nabla \bar u + \nabla(\eta\nabla \bar u)
\end{equation}
\begin{equation}
    \dot \eta = \alpha \cdot \nu\nabla^2\eta - \beta \cdot \bar u\nabla\eta
\end{equation}
\begin{equation}
    \alpha, \beta = f_\theta(\nabla \bar u, \nabla \eta, \nu, \delta x; x)
\end{equation}
\begin{equation}
    \eta_0 = g_{\theta}(\bar u_0, \nu; x).
\end{equation}

Model \textit{transport-I-p} is nearly identical to \textit{transport-I}, however the $\eta$ transport equation has an additional production term $\gamma$:
\begin{equation}
    \dot{\bar{u}} = \nu\nabla^2\bar u - \bar u\nabla \bar u + \nabla\eta
\end{equation}
\begin{equation}
    \dot \eta = \alpha \cdot \nu\nabla^2\eta - \beta \cdot \bar u\nabla\eta + \gamma
\end{equation}
\begin{equation}
    \alpha, \beta, \gamma = f_\theta(\nabla \bar u, \nabla \eta, \nu; x)
\end{equation}
\begin{equation}
    \eta_0 = g_{\theta}(\bar u_0, \nu; x).
\end{equation}

Model \textit{transport-II-p} is equivalent to \textit{transport-II} except for the additional production term:
\begin{equation}
    \dot{\bar{u}} = \nu\nabla^2\bar u - \bar u\nabla \bar u + \nabla(\eta\nabla \bar u)
\end{equation}
\begin{equation}
    \dot \eta = \alpha \cdot \nu\nabla^2\eta - \beta \cdot \bar u\nabla\eta + \gamma
\end{equation}
\begin{equation}
    \alpha, \beta, \gamma = f_\theta(\nabla \bar u, \nabla \eta, \nu; x)
\end{equation}
\begin{equation}
    \eta_0 = g_{\theta}(\bar u_0, \nu; x).
\end{equation}

The last model, \textit{smag-ml}, attempts to improve on the standard constant Smagorinsky model by learning a spatially varying Smagorinsky constant, computed via neural network:
\begin{equation}
    \dot{\bar{u}} = \nu\nabla^2\bar u - \bar u\nabla \bar u + \nabla(\nu_t\nabla \bar u)
\end{equation}
\begin{equation}
    \nu_t = (\eta\delta x)^2|\nabla \bar u|
\end{equation}
\begin{equation}
    \eta = f_\theta(\nabla \bar u, \nu, \delta x; x).
\end{equation}

All of the models used in our analysis are tabulated in \autoref{tab:models} of the appendix.

\subsection{Neural network architecture}
To leverage our choice of pseudospectral discretization, we use Fourier Neural Operators (FNO) as our network architecture \cite{FNO}. All functions with subscript $\theta$ in the previous section correspond to an FNO architecture. FNOs combine pointwise in real space and pointwise in Fourier space linear operations with pointwise nonlinearities in real space to enable learning of a large suite of resolution-invariant function classes, including linear and nonlinear operators.

\begin{figure*}
  \centering
  \includegraphics[width=0.8\linewidth]{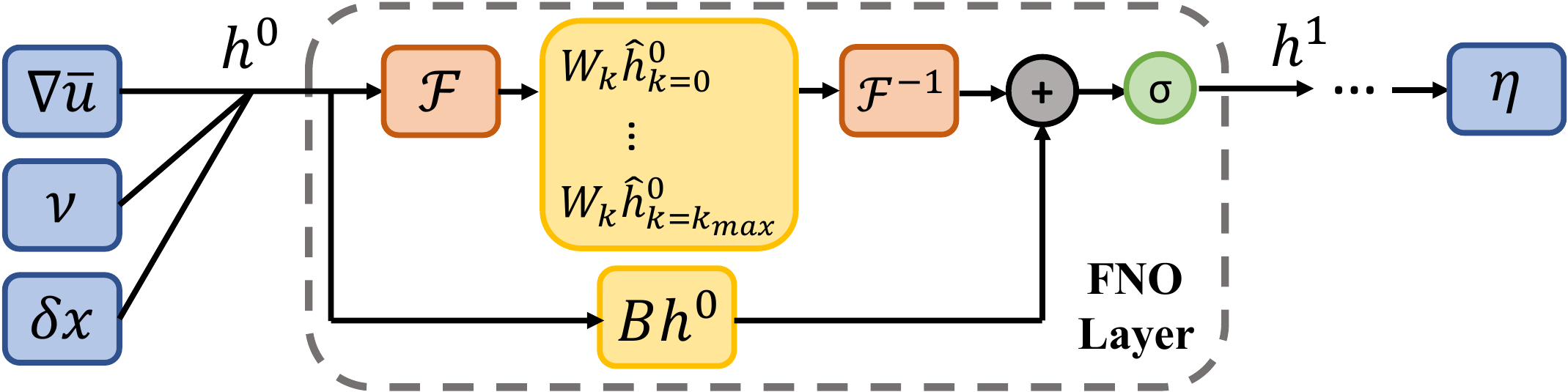}
  \caption{An example schematic of the FNO neural network architecture for the function $f_\theta$ in \textit{smag-ml}. At each time step, the network takes as input $\nabla \bar u$, $\nu$, and $\delta x$, computing the $\eta$ field that corresponds to a spatially varying subgrid term. Within the network, there are two FNO layers that include a local linear bias from matrix $B$ and a transformation of the Fourier modes $\hat h_k$ with matrix $W_k$. These are summed and passed through a non-linearity to the next layer.}
  \label{fig:fno}
\end{figure*}

Given an input feature $h^n \in \mathbb{R}^{N\times c_{in}}$, where $c_{in}$ is the number of channels or scalar fields, an FNO layer is defined as:
\begin{equation}
    h^{n+1} = \sigma(Bh^n + \mathcal{F}^{-1}(W_k\mathcal{F}(h^n)_k)), 
    \label{eqn:fno}
\end{equation}
where $B \in \mathbb{R}^{c_{in} \times c_{out}}$ is a 2-dimensional tensor that is contracted along $c_{in}$ identically pointwise in real space and $W \in \mathbb{R}^{k_{max} \times c_{in} \times c_{out}}$ is a 3-dimensional tensor that is contracted along $c_{in}$ pointwise in Fourier space for each wavemode $k$ up to $k_{max}$. Therefore, the Fourier transform of the input is truncated at $k_{max}$. The output of these operations are summed and a pointwise nonlinearity $\sigma$ is applied. The parameters of the layer are given by the tensors $W$ and $B$. \autoref{fig:fno} illustrates the algorithm for $\eta = f_\theta(\nabla \bar u, \nu, \delta x; x)$ in the \textit{smag-ml} model.

\subsection{Optimization}
Network parameters are optimized with respect to the loss function:
\begin{equation}
    L = \sum_{i=1}^{T}\sum_{j=1}^{M} \frac{(\bar u_{model}(x_j,t_i)-\bar{u}_{DNS}(x_j,t_i))^2}{M*T}
\end{equation}
\begin{equation}
\hat{\bar{u}}_{DNS}=G\star \hat{u}_{DNS},
\end{equation}
where $u_{DNS}$ is the ground truth velocity field resolved on a grid of $N$ points, $\bar{u}_{DNS}$ is filtered using sharp low-pass filter $G$ with cutoff $k=M/2$, $x_j$ correspond to the $M$ collocation points on the coarse grid, and $t_i$ correspond to the $T$ timesteps in the discretized solution trajectory.

\begin{figure*}
  \centering
  \includegraphics[width=0.8\linewidth]{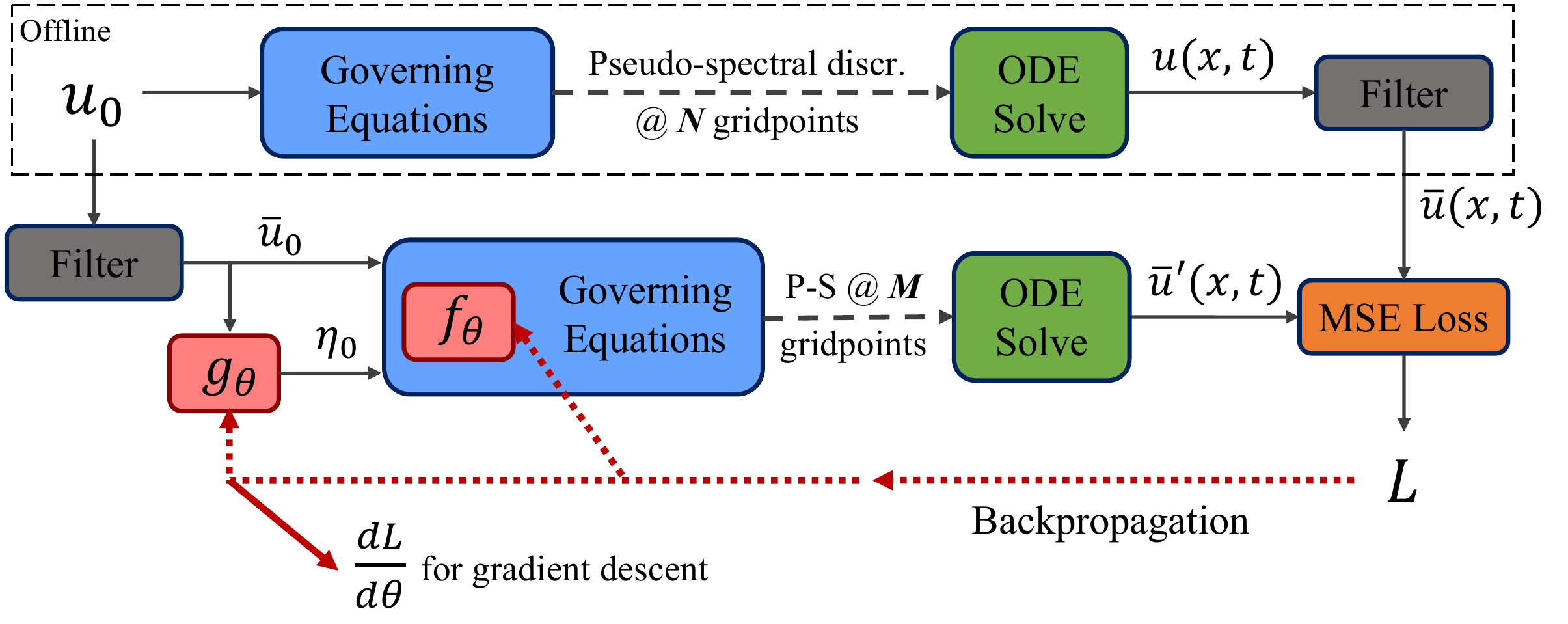}
  \caption{A schematic of the training procedure for learning the parameters in the data-driven models. An initial velocity field is evolved (offline) according to the governing equations on a DNS-resolving grid and filtered to generate ground truth data. During the optimization loop, the same field is filtered and evolved with the data-driven model to compute an estimate for the filtered field. The gradients are backpropagated through the ODE solve and spatial discretization to update the parameters.}
  \label{fig:schematic}
\end{figure*}

Optimization is achieved with the gradient-based optimizer ADAM, shown to be effective for deep learning applications \cite{ADAM}.
Gradients are computed with the automatic differentiation (AD) platform available in \textit{Pytorch} via reverse-mode autodifferentiation, or backpropagation \cite{pytorch}. We show a schematic of the training procedure in \autoref{fig:schematic}.

Most of the operations in our solution algorithm consist of basic linear operations and pointwise nonlinearities, operations that have formed the foundation of classical multi-layer perceptron machine learning architectures. Deep learning libraries have readily supported backpropagation through these operations for many years and thus we do not discuss implementation details further. However, we highlight two slightly more complex components of the algorithm corresponding to aspects of the spatial and temporal discretization and discuss their gradient computation.

\subsubsection{Gradients of the Fourier transform}
In reverse-mode gradient accumulation \cite{speelpenning}, we wish to compute vector-Jacobian products (VJPs) \cite{pytorchAD,griewank} of each function $f$ in the algorithm:
\begin{equation}
    \text{VJP} = \mathbf{v}^T f'(\mathbf{x}),
\end{equation}
given the cotangent vector $\mathbf{v}$, input $\mathbf{x}$, and Jacobian $f'(\mathbf{x})$. The VJP can then be passed to the previous function in the computational graph to compute the next VJP.

We leverage spectral basis expansion to compute spatial derivatives in the flow and inside the FNO layers. This is implemented with a Discrete Fourier Transform (DFT) operation. Since the DFT is ultimately a linear operation
\begin{equation}
    \mathcal{F}_{DFT}(\mathbf{x}) = \mathbf{F}\mathbf{x},
\end{equation}
with DFT matrix $\mathbf{F}$, the Jacobian of the transform is simply the matrix $\mathbf{F}$. To compute the VJP, we can equivalently perform matrix-vector multiplication with the adjoint operator $\mathbf{F^*}$. Since $\mathbf{F^*}=N\mathbf{F}^{-1}$, we can compute the VJP as an inverse Fourier transform multiplied by the size of the signal. The benefit of this approach comes from the ability to implement Fast Fourier Transform \cite{fftw} (FFT) algorithms during the backwards pass as well, instead of constructing the entire DFT matrix explicitly.

\subsubsection{Propagating gradients through time integration}
Time evolution of the velocity field is accomplished by solving the ODE resulting from  spatial discretization of the field using an arbitrary integration method. A naive implementation of AD would propagate gradients through each of basic operations in the solver. Depending on the solver tolerance, order, and trajectory length, this may result in a very large number of operations, and all intermediate steps must be saved in memory for AD to work. In our experiments, backpropagating through even moderately sized trajectory lengths, networks, and batch sizes could exceed the 16GBs of memory on our NVIDIA V100 GPU using naive autograd.

Therefore, a more memory efficient approach is desired. We use the adjoint method in time, which has been proposed for use in machine learning applications by Chen et. al \cite{Chen}.
The forward ODE system is given by the evolution of the discrete Fourier modes:
\begin{equation}
    \frac{d\hat{\mathbf{u}}_k(t)}{dt} = f(\hat{\mathbf{u}}_k(t),\theta),
\end{equation}
where $\theta$ represents any learnable parameters in the model. The loss can be written as an integral:
\begin{equation}
    L = \int_{t_0}^{t_T} l(\hat{\mathbf{u}}_k(t)) dt  .
\end{equation}
The adjoint method produces a second ODE system that evolves the adjoint variable $\lambda$ backwards in time, given by the equation:
\begin{equation}
    \frac{d\mathbf{\lambda}(t)}{dt} = -\lambda(t)^T \frac{\partial f(\hat{\mathbf{u}}_k(t),\theta)}{\partial \hat{\mathbf{u}}_k} + \frac{\partial l(\hat{\mathbf{u}}_k(t))}{\partial \hat{\mathbf{u}}_k},
\end{equation}
and an integral:
\begin{equation}
    \frac{dL}{d\theta} = - \int_{t_T}^{t_0} \lambda(t)^T \frac{\partial f(\hat{\mathbf{u}}_k(t),\theta)}{\partial \theta} dt,
\end{equation}
to compute the parameter sensitivities. The RHS of the equations involve VJPs that are efficiently evaluated with AD.
The adjoint system can again be solved with an arbitrary solver, potentially even different than the forward solver.
Specifically, we use the interpolated adjoint method, which reduces the memory footprint by constructing an interpolation of the forward solution, which appears in the adjoint equations. During the forward pass, the solution is saved at a selected few timesteps and reused in the adjoint computation via interpolation during the backwards pass.

\section{Experiments and results}
\label{sec:experiment}
To assess the quality and practicality of our proposed models, we seek to study model performance across a range of Burgers solutions. 

\subsection{Viscosity generalization and temporal stability}
%\textcolor{red}{Romit: Could we call it temporal stability - extrapolation is a bit of a touchy term.}

In this experiment, we examine model generalizability with respect to initial conditions of the velocity field and the viscosity parameter $\nu$. In addition, we investigate stability of the models by extrapolating in time relative to the training data.

The models are trained using the optimization procedure described in the previous section on a ground truth dataset of 100 Burgers solutions. We vary the initial conditions of the velocity field by randomizing the Fourier modes of the initial field. The $k$th Fourier mode is denoted by $\hat u_k$. The dataset is generated with the parameters tabulated in \autoref{tab:model_params}.

\renewcommand{\arraystretch}{1.5}
\begin{table}[h]
\centering
\begin{tabularx}{0.7\linewidth}{cm{1pt}|m{2pt}X} 
\textbf{Parameter} &&& \textbf{Range} \\ \hline
$\nu$       &&&  $[10^{-4},10^{-3})$\\
$\hat u_k$  &&&  $\left([-0.5,0.5)+[-0.5,0.5)i\right)e^{-0.5k}$, $k=0..32$\\
$t$         &&&  $[0,3)$\\
$N$         &&&  4096\\    
$M$         &&&  64\\ 
\end{tabularx}
\caption{Parameters and their distributions used to generate the training set for the first experiment, including the viscosity $\nu$, the Fourier modes of the initial velocity field $\hat u_k$, the timespan for integration $t$, the DNS grid resolution $N$, and the coarse grid resolution $M$.}
\label{tab:model_params}
\end{table}

The neural networks in the trainable models contain 2 FNO layers with $k_{max}=16$, 128 hidden channels, and appropriate input and output sizes. The first layer uses a ReLU nonlinearity while the second uses the identity. The models are trained for 500 epochs using the ADAM optimizer with learning rate varying from $10^{-3}$ to $10^{-1}$ depending on the model. Generally, models such as \textit{smag-ml} that are more constrained by inductive biases required larger learning rates.

Model performance is gauged using the mean-squared error of the predictions relative to the ground truth solution on an unseen test set. We leverage two test sets to assess generalizability across viscosities and stability in time. The first test set consists of $25 \times 25$ solutions -- 25 discrete viscosities in the range of $10^{-4.5}$ -- $10^{-2}$ and 25 trajectories, or different initial conditions, per viscosity. We compute the model error for each prediction and plot the average ensemble error as a function of viscosity.

The second test set contains 25 trajectories at $\nu=10^{-3}$. Here, we evolve the trajectories between $t=[0,10)$, over 3 times longer than the training set, to see how the models behave outside of the training regime. The average error across all trajectories is plotted as a function of time.

\begin{figure*}
  \centering
  \includegraphics[width=\linewidth]{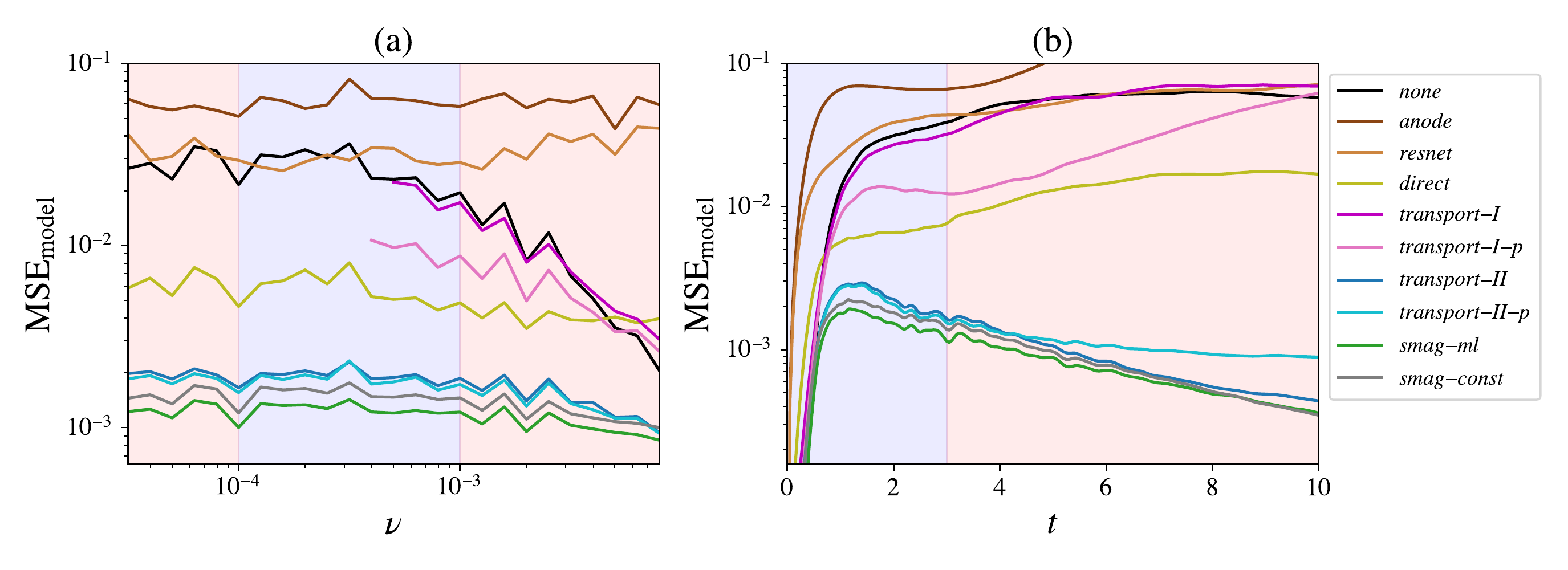}
  \caption{Blue shaded regions indicate the section of the plot corresponding to the training regime, while red shading indicates a test regime. (a) Plot of the ensemble mean-squared error relative to the ground truth DNS for each model as a function of viscosity. All models generally improve over the no closure baseline except $anode$, but only $smagorinsky$ improves of the constant Smagorinsky model. (b) Ensemble MSE for each model as a function of time. The inclusion of a production term gives rise to anomalous behavior in the extrapolation region.}
  \label{fig:all_models}
\end{figure*}

Two non-trainable models \textit{none} and \textit{smag-const} provide some bounding benchmarks for model comparison. In \autoref{fig:all_models}(a), at the upper end, the \textit{none} model with no closure results in high error at low viscosities, where the dissipative length scales are small, with progressively lower error at higher viscosities where the dissipative length scales become larger. An effective closure model should at the very least perform better than this, otherwise no additional physics have been learned to improve over the baseline governing equations solved on a coarse grid. At the lower end, the \textit{smag-const} model is a well-known closure model that has been shown to perform well in a wide range of viscosities. An ideal model would produce lower loss than this baseline, indicating that the neural closure functional form can show improvements over classical closure models from theory.

Model performance is delineated well by model accuracy relative to the baselines in \autoref{fig:all_models}(a). Model \textit{anode} has the highest loss, greater than the \textit{none} baseline, thus showing no improvement over the governing equations. Given that the \textit{anode} model contains no inherent physics or structural form to the PDE, this result is unsurprising as the model must learn both the governing dynamics and any closure approximations. Despite \textit{resnet} having a similar architecture to \textit{anode}, \textit{resnet} can achieve marginally better accuracy, although it is still no better than the \textit{none} baseline. Theoretically, \textit{anode} should be able to learn the same optima as \textit{resnet}, since \textit{anode} has the same functional form as \textit{resnet}, except solved using an adaptive time-stepper. However, actually reaching those optima can be challenging during optimization due to the ODE inductive bias.

Models \textit{transport-I} and \textit{transport-I-p} show marginal improvement over \textit{none}, with \textit{transport-I-p} the better of the two. These models were particularly difficult to train, producing very stiff ODEs that made integration difficult. This is apparent in the fact that some of the viscosities are not shown as the stiffness of those systems led to intractable solutions. We hypothesize that this could be due to overfitting the training set, since all of the training samples could be solved in a reasonable amount of time.

Model \textit{direct} lies firmly between the two benchmarks. Accuracy is relatively uniform with little dependency on viscosity. Thus, while \textit{direct} consistently improves over \textit{none} at low viscosities, the advantage decreases with increasing viscosity, becoming a disadvantage at the highest viscosities.

Models \textit{transport-II} and \textit{transport-II-p} have accuracies comparable to \textit{smag-const} and perform almost identically to each other within the viscosity test. Generally, we see a consistent positive bias over the baseline closure benchmark that diminishes only at the highest viscosities.

Model \textit{smag-ml} is the only model with lower error than \textit{smag-const} everywhere in \autoref{fig:all_models}(a). Given that the model leverages the functional form of the classical Smagorinsky model and only adds a spatial dependency, again this result is perhaps unsurprising. However, it is promising that the improvements persist beyond the training region.

Overall, besides models \textit{none}, \textit{transport-I}, and \textit{transport-I-p}, accuracy is mostly independent of the viscosity within this range. While models have been trained on viscosities between $10^{-4}$ and $10^{-3}$, they can be used on viscosities well outside this range, with little to no erroneous behavior outside the training regime.

\autoref{fig:all_models}(b) shows model accuracy on the second test set. The blue region indicates the time span the model was trained on, while the red region shows the extrapolation range. The model hierarchy by accuracy is generally preserved here, i.e. \textit{anode} performs the worst, \textit{smag-ml} the best, and \textit{transport-II} and \textit{transport-II-p} are comparable to \textit{smag-const}. 

However, the stability test highlights the effect of the production term in the models. The models largely have a monotonically decreasing slope in the error over time, except for \textit{anode}, \textit{transport-I-p}, and \textit{transport-II-p}, which display aberrant behavior in the extrapolation range. The inclusion of the production term results in improvements within the training time regime, and in the case of \textit{transport-I} and \textit{transport-I-p}, the difference is significant. Outside of this range, the production term can hinder the solution and lead to growing errors. Deviations between \textit{transport-II} and \textit{transport-II-p} can be seen at $t>5$, and error from \textit{transport-I-p} grows to match \textit{transport-I} at $t=10$.

\begin{table}[h]
\centering
\begin{tabularx}{0.6\linewidth}{rm{2pt}|m{2pt}X}
\textbf{Model} &&& \textbf{Relative evaluation time} \\ \hline
\textit{none}               &&& 1.0 \\
\textit{anode}              &&& 2.54 \\
\textit{resnet}        &&& 0.37 \\
\textit{direct}             &&& 2.97 \\
\textit{transport-I}          &&& 3.36 \\
\textit{transport-I-p}         &&& 3.40 \\
\textit{transport-II}         &&& 3.48 \\
\textit{transport-II-p}        &&& 3.52 \\
\textit{smag-ml}        &&& 3.24 \\
\textit{smag-const}  &&& 1.36 \\
\end{tabularx}
\caption{Table of evaluation times for the models on the second test set. Times are normalized by the \textit{none} model, which is the cheapest differential equation-based model to compute.}
\label{tab:model_evals}
\end{table}

\autoref{tab:model_evals} shows the model evaluation times relative to \textit{none}. All of the models except for \textit{resnet} are more expensive than this baseline, which is expected considering that the models require more operations and Fourier transforms to compute than the governing dynamics. We found that the model evaluation times have limited correlation with the number of operations needed at each time step, and are more heavily influenced by the stiffness of the resulting ODE system, given the adaptive integration scheme. Thus, \textit{resnet} is much more efficient than other models because it requires the fewest function evaluations, coinciding with the fixed Euler time steps.

\begin{figure*}
  \centering
  \includegraphics[width=\linewidth]{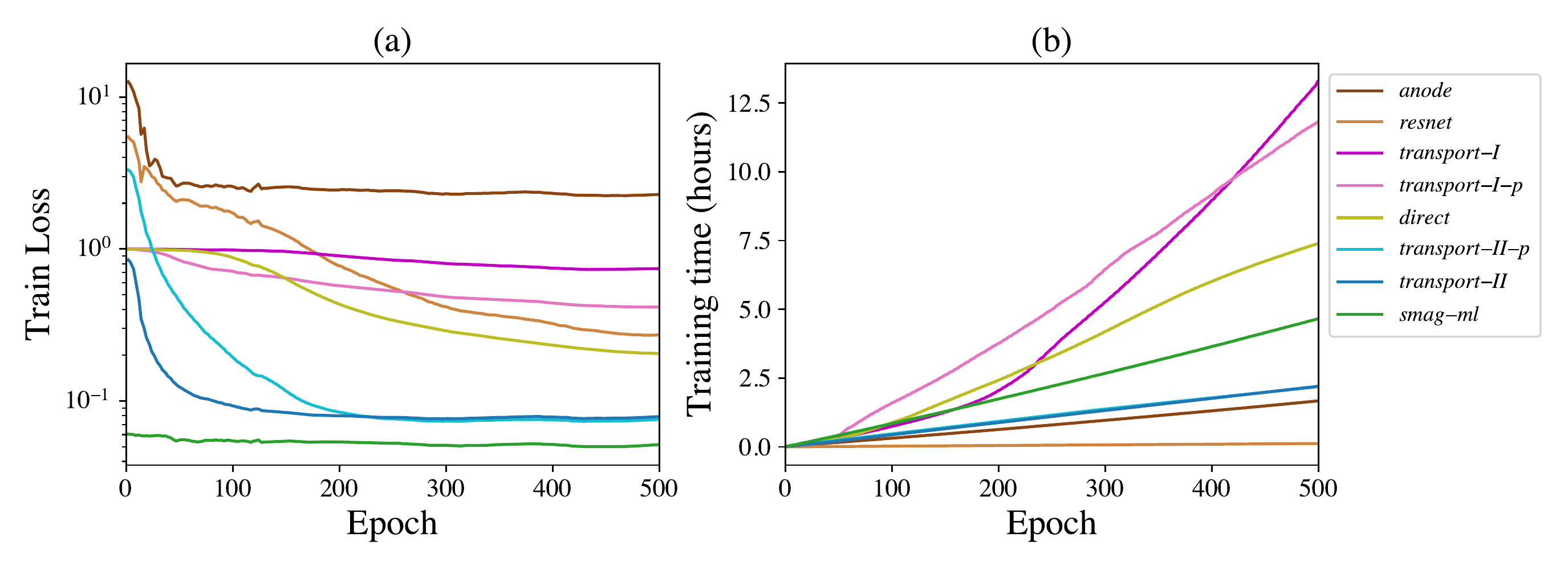}
  \caption{Training curves for the models in the first experiment. (a) MSE loss of the models, normalized by \textit{none}, as a function of epoch. There is significant variability between the models in time to convergence and magnitude of the overall change in loss. (b) Training time as a function of epoch. For some models, the computational cost of evaluation remains constant, leading to linear behavior. For others, the computational cost increases with training iteration, demonstrating non-linear behavior with a positive second derivative.}
  \label{fig:model_training}
\end{figure*}

We also show training curves for this experiment in \autoref{fig:model_training}, including the train loss and training time as a function of epoch. 
We point out the train loss of \textit{resnet} in \autoref{fig:model_training}(a), which is vastly different than the test loss in \autoref{fig:all_models}, different behavior from the other models tested. We believe the use of differential equation inductive biases in the other models close the learning gap between training and testing data distributions, promoting data-efficient and generalizable models.
\autoref{fig:model_training}(b) clearly demonstrates the effect of ODE stiffness. Since the number of operations per function evaluation in each model is fixed, one might expect the training time to be linear with epoch. However, as models like \textit{transport-I} and \textit{direct} evolve in training, the curves become steeper, indicating more function evaluations required for time integration.

\subsection{Resolution invariance}
In the previous experiment, we have shown that many of the models generalize well over a wide range of viscosities and time. In this experiment, we wish to examine our approach with regards to the coarse-grain scale. In the previous experiment, we fixed $M=64$, here we will vary $M$ and thus the cutoff of the low-pass filter, $k=M/2$.

Again, we generate a training dataset of 100 Burgers solutions. Since we desire for all the cutoffs $M/2$ to lie in the inertial region of the flow spectrum, we slightly modify the distribution of the initial condition, the time span for integration, and the ground truth grid density $N$, realizing a wider inertial range. We divide the set into 4 batches with $M=64,128,256,512$. The dataset parameters are tabulated in \autoref{tab:res_params}.

\begin{table}[h]
\centering
\begin{tabularx}{0.7\linewidth}{cm{1pt}|m{2pt}X} 
\textbf{Parameter} &&& \textbf{Range} \\ \hline
$\nu$       &&&  $[10^{-4},10^{-3})$\\
$\hat u_k$  &&&  $[-0.05,0.05)+[-0.05,0.05)i$, $k=0..32$\\
$t$         &&&  $[0,6)$\\
$N$         &&&  8192\\    
$M$         &&&  64, 128, 256, 512\\ 
\end{tabularx}
\caption{Parameters and their distributions used to generate the training set for the second experiment, including the viscosity $\nu$, the Fourier modes of the initial velocity field $\hat u_k$, the timespan for integration $t$, the DNS grid resolution $N$, and the coarse grid resolutions $M$.}
\label{tab:res_params}
\end{table}

We train the two best performing models from the previous experiment, \textit{transport-II} and \textit{smag-ml}, as well as two modifications to the \textit{smag-ml} model to understand the impact of certain neural network architectural choices. These models are trained for 200 epochs, which was sufficient for convergence, again using the ADAM optimizer with learning rate varying from $10^{-3}$ to $10^{-1}$ depending on the model.

Specifically, we add two models, \textit{smag-ml-small} and \textit{smag-ml-local}. \textit{smag-ml-small} reduces the number of hidden channels in the network to 2 from 128. Since the FNO architecture requires each of the channels to be Fourier transformed and inverse transformed at each timestep, the size of the hidden layer has a large impact on the computational efficiency of the model. Therefore, we wish to see how small the model can be made without sacrificing accuracy. \textit{smag-ml-local} switches the FNO layers for standard linear layers resulting in a purely local closure correction. The FNO is a non-local operator given the pointwise multiplication in Fourier space. Replacement with a linear layer gives only pointwise multiplication by a linear operator in real space. Thus, we can compare the efficacy of a non-local functional form with a local functional form.

To test the models, we generate a new test set that includes variation in $M$. Here, we look at a set of $13 \times 25$ trajectories -- 13 discrete $M$'s in the range of $2^6$-$2^9$, including interpolation between the training discretizations, and 25 samples per $M$. The $\nu$ and $\hat u_k$ values are drawn from the same distributions as the training set, and $t=[0,20)$.

\begin{figure*}
  \centering
  \includegraphics[width=0.9\linewidth]{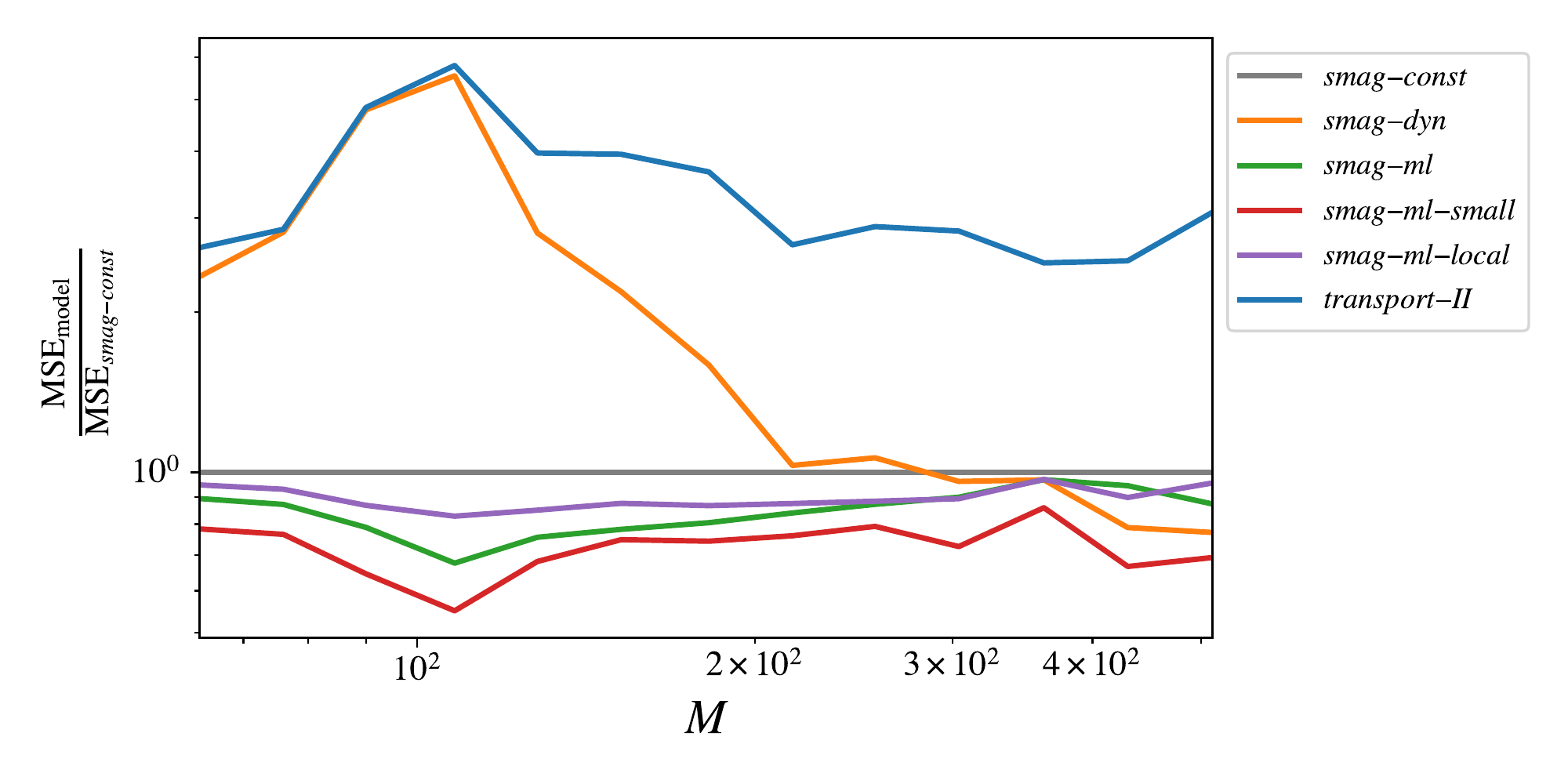}
  \caption{The ensemble mean-squared error of various models normalized by the MSE of the constant Smagorinsky model, one of the baseline we hope to improve over. The adaptive ML Smagorinsky models can reduce the error up to 40\%, with \textit{smag-ml-small} demonstrating the best performance. \textit{smag-dyn} can only achieve lower error than \textit{smag-const} at high resolutions, where the inertial region is more fully resolved.}
  \label{fig:resolution}
\end{figure*}

\begin{figure}
  \centering
  \includegraphics[width=0.9\linewidth]{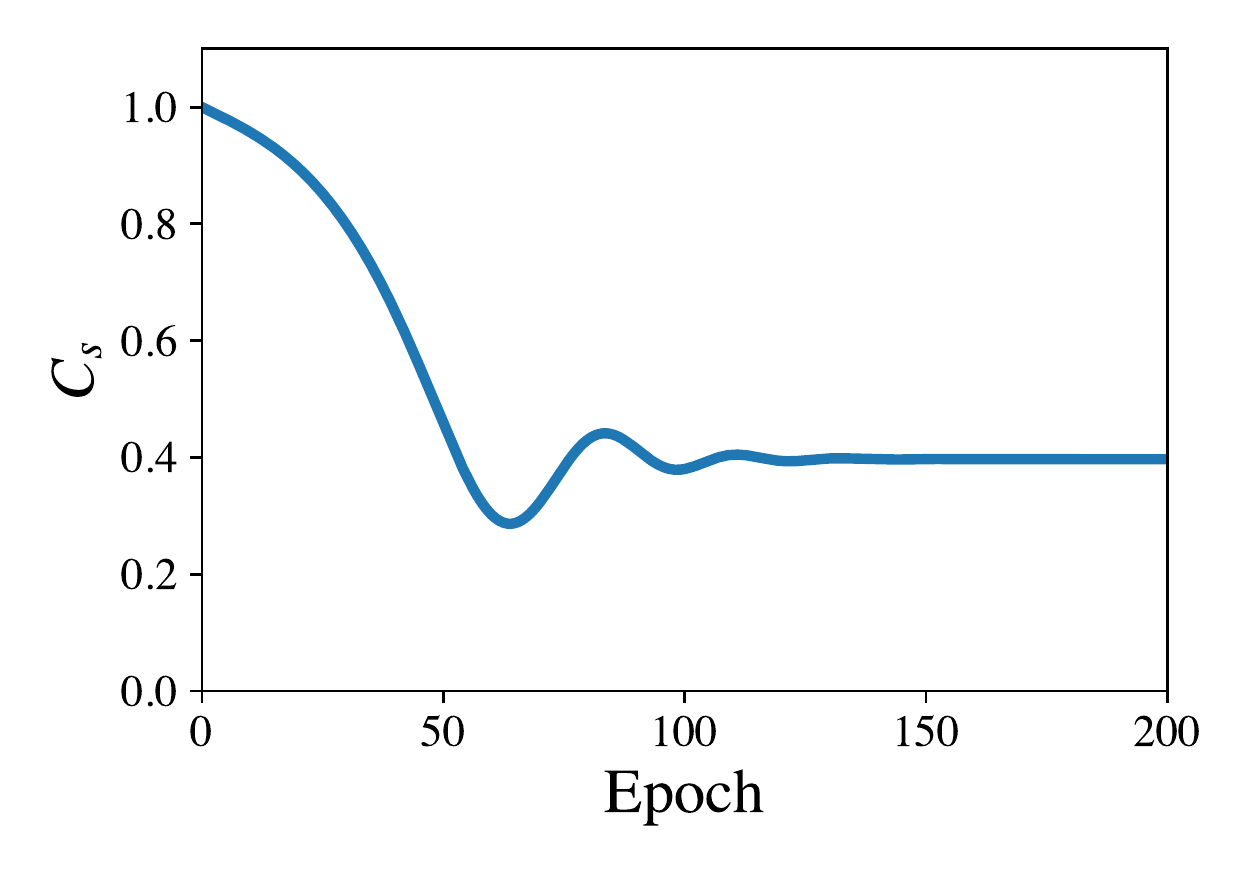}
  \caption{Convergence of the Smagorinsky constant $C_s$ to 0.4 as optimized by gradient descent on the training dataset.}
  \label{fig:Cs}
\end{figure}

\autoref{fig:resolution} shows the MSE of the models normalized by the constant Smagorinsky baseline. Since the training dataset encompasses variations in both $\nu$ and $M$, we train the Smagorinsky constant via gradient descent on the same training dataset as the models to obtain the optimal constant instead of hand-tuning. The training procedure converged clearly to a value of $C_s=0.4$ as shown in \autoref{fig:Cs}. We also include a comparison to the non-trainable \textit{smag-dyn} baseline, given that most of the models in this experiment are Smagorinsky model derivatives.

Similar to the previous experiment, \autoref{fig:resolution} shows that model \textit{transport-II} results in greater error than \textit{smag-const}, while the adaptive Smagorinsky approaches can improve over the constant model. Even when the ideal $C_s$ is fit to the data, the spatiotemporally varying models outperform the baseline. Within the adaptive trainable Smagorinsky models, the local correction can improve over the constant baseline at all resolutions, though not as much as the other two models. The two non-local models generally perform better than all others, which fits with our understanding that turbulence is a non-local phenomenon. However, surprisingly, the \textit{smag-ml-small} model ultimately shows a universal advantage over both the dynamic and constant Smagorinsky baselines. Comparing the training and validation losses of the models provides some insight. The overparameterized \textit{smag-ml} model overfits the training set resulting in worse performance on the test set. While this may be ameliorated by a larger training dataset, we do not pursue this route and leave this study for future work. Instead, we show that we can learn an effective and computationally efficient correction to the baseline Smagorinsky model with limited training data, due to our differentiable physics approach to end-to-end training of ML models.

\begin{table}[h]
\centering
\begin{tabularx}{0.6\linewidth}{rm{2pt}|m{2pt}X}
\textbf{Model} &&& \textbf{Relative evaluation time} \\ \hline
\textit{none}               &&& 1.0 \\
\textit{smag-const}  &&& 1.33 \\
\textit{smag-dyn}    &&& 2.35 \\
\textit{smag-ml}        &&& 3.65 \\
\textit{smag-ml-small}  &&& 3.47 \\
\textit{smag-ml-local}  &&& 1.77 \\
\textit{transport-II}         &&& 3.97 \\
\end{tabularx}
\caption{Table of evaluation times for the models used in the resolution experiment. Times are normalized by the \textit{none} model, which is the cheapest to compute.}
\label{tab:res_evals}
\end{table}

\autoref{tab:res_evals} lists the normalized evaluation times for the models in this experiment. While the FNO-based models are unequivocally more expensive than any of the baselines, \textit{smag-ml-local} provides an interesting compromise between computational cost and accuracy, achieving lower error than the baselines everywhere except for the highest resolutions, where \textit{smag-dyn} improves over \textit{smag-const}.

\begin{figure*}
  \centering
  \includegraphics[width=\linewidth]{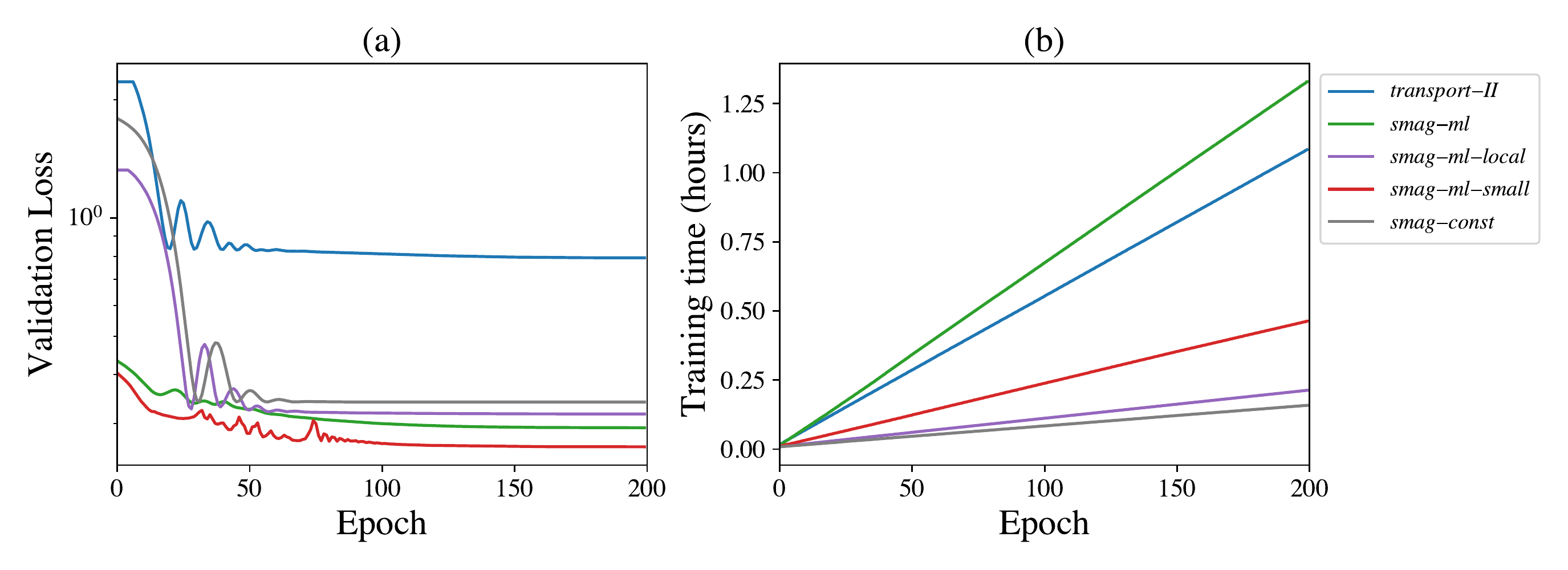}
  \caption{Training curves for the models in the second experiment. (a) MSE loss of the models, normalized by \textit{none}, as a function of epoch. The inductive biases of these models lead to fast convergence. (b) Training time as a function of epoch. Although \textit{smag-ml} and \textit{smag-ml-small} have comparable final evaluation costs, the differences become significant during training, in terms of overall optimization time.}
  \label{fig:resolution_training}
\end{figure*}

The training curves for the models are shown in \autoref{fig:resolution_training}. From \autoref{fig:resolution_training}(a), convergence was achieved in 200 epochs and from \autoref{fig:resolution_training}(b), these models did not display the same kind of non-linear behavior in the training time. \textit{smag-ml-small} was much faster to train than \textit{smag-ml}, despite the comparable evaluation times.

\begin{figure*}
  \centering
  \includegraphics[width=\linewidth]{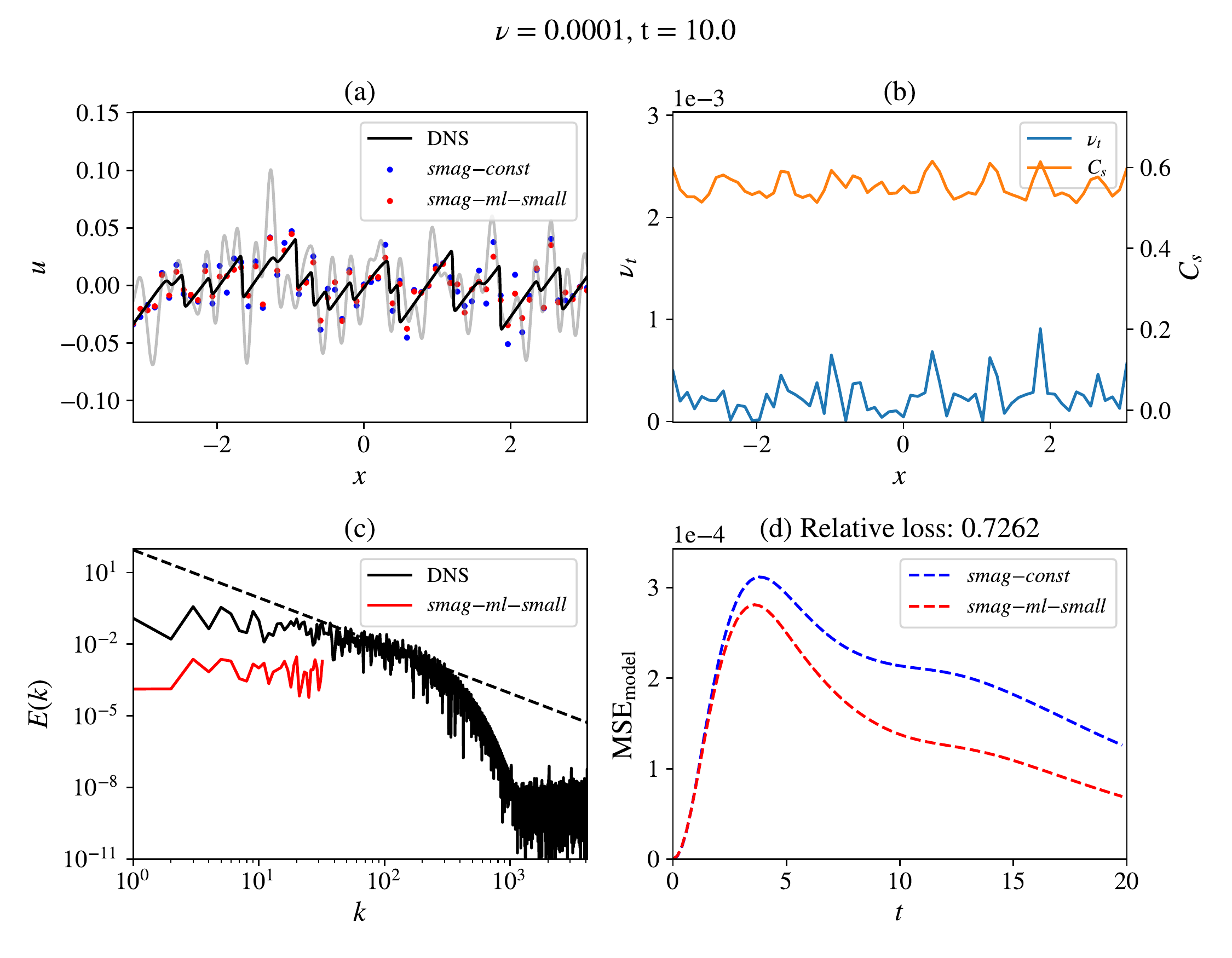}
  \caption{Statistics and snapshots comparing results from closure model \textit{smag-ml-small} to a baseline and DNS with $M=64$. (a) Sample snapshot of the velocity field from DNS, the \textit{smag-const} baseline, and \textit{smag-ml-small}. (b) Sample snapshot of the closure model eddy viscosity field and spatially varying Smagorinsky constant from \textit{smag-ml-small}. (c) Snapshot of the energy spectra of the sample from DNS and  \textit{smag-ml-small}. The cutoff $M/2$ lies in the very upper region of the inertial range. (d) MSE of the model and baseline over time.}
  \label{fig:64}
\end{figure*}

\begin{figure*}
  \centering
  \includegraphics[width=\linewidth]{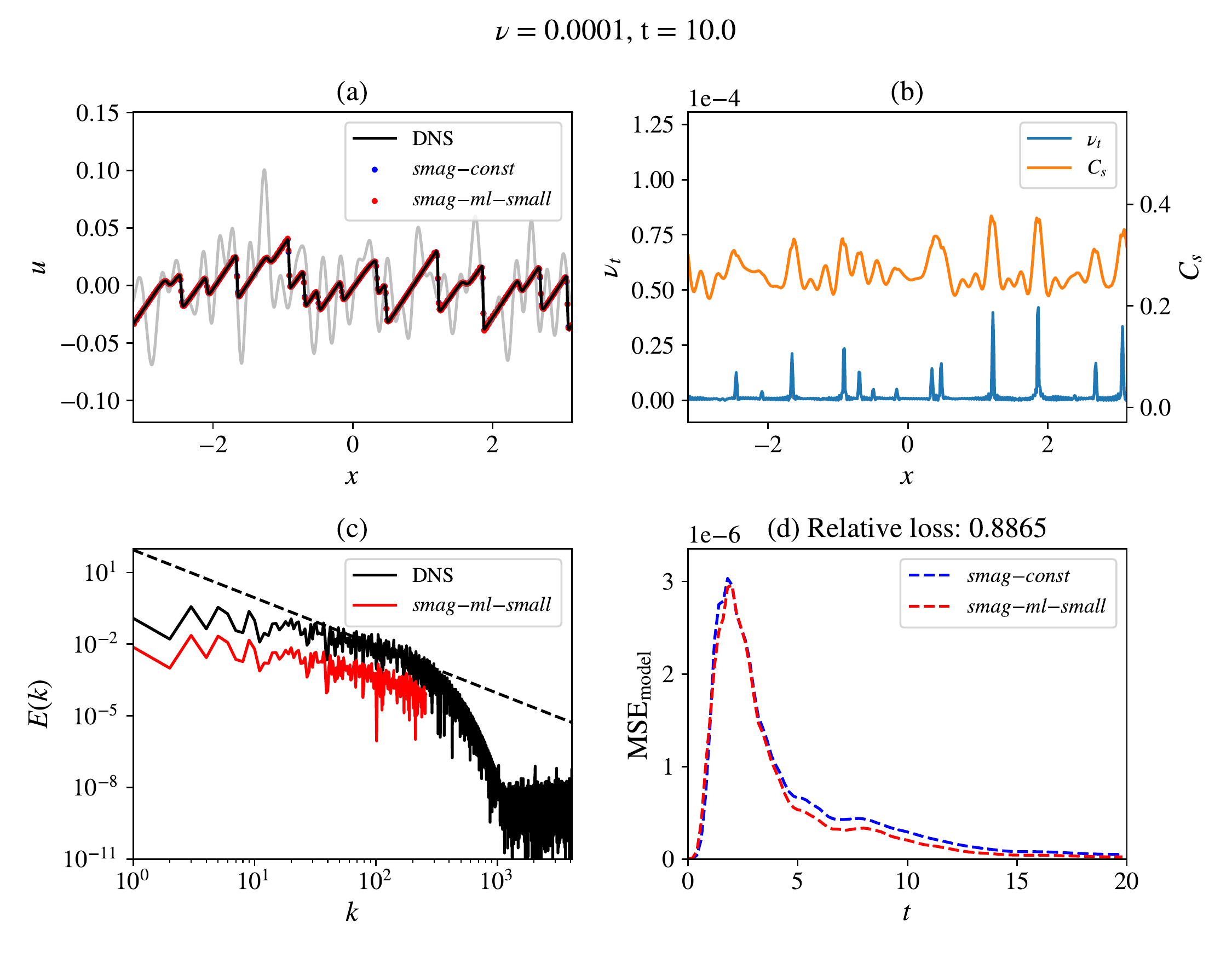}
  \caption{Statistics and snapshots comparing results from closure model \textit{smag-ml-small} to a baseline and DNS with $M=512$. (a) Sample snapshot of the velocity field from DNS, the \textit{smag-const} baseline, and \textit{smag-ml-small}. (b) Sample snapshot of the closure model eddy viscosity field and spatially varying Smagorinsky constant from \textit{smag-ml-small}. (c) Snapshot of the energy spectra of the sample from DNS and  \textit{smag-ml-small}. The cutoff $M/2$ lies very close to the dissipative range. (d) MSE of the model and baseline over time.}
  \label{fig:512}
\end{figure*}

Lastly, we show some examples of predictions from model \textit{smag-ml-small} compared to \textit{smag-const}. \autoref{fig:64} displays results from $M=64$, where the cutoff wavenumber is close to the energy containing region of the spectra. \autoref{fig:512} presents results from $M=512$, where the cutoff wavenumber is on the other side of the inertial region close to the dissipative region. The snapshots of the velocity field in (a) and the Smagorinsky constant and eddy viscosity in (b) are taken at $t=10$. (c) compares the DNS and model energy spectra at $t=10$, where the cutoff wavenumber is indicated by the model spectrum. (d) illustrates the MSE loss of the model \textit{smag-ml-small} and \textit{smag-const} over time, including the relative loss of the model to the constant Smagorinsky baseline.

\section{Conclusions}
\label{sec:concl}
In this work, we systematically investigate the efficacy of various machine learning based closure models for Burgers turbulence using an end-to-end learning framework. We aim to develop a better understanding of how the functional form of the data-driven closure impacts its ability to accurately model subgrid scales, and seek to design a data-driven closure that is broadly applicable and ideally outperforms baseline closure models. Our models are discretization independent and thus defined by their PDE form and neural network hyperparameters. Models are trained in an end-to-end fashion, meaning the entire closure PDE is solved with a pseudo-spectral method at each iteration of the optimization procedure. Differentiable programming closes the gap between typical \textit{a priori} learning and separate \textit{a posteriori} testing that is common in ML closure modeling today. Given a differentiable solution algorithm, models can be trained via gradient descent directly on an \textit{a posteriori} loss function. This approach allows for great flexibility in model design. Just as neural network hyperparameters can be easily tuned and adjusted, the PDE form of the closure can be modified to incorporate various degrees of known physics or physical assumptions. Once the high level PDE has been defined, the system is solved and evaluated on the loss function, the MSE of the velocity field relative to a ground truth DNS simulation.

We asses the models on a wide range of Burgers systems to determine their generalizability. Given the spread of Reynolds numbers encountered in flow problems and the need for a closure model that can accommodate this range, we test our models' performance on unseen Burgers systems with viscosities spanning 2.5 orders of magnitude, including interpolation of the training viscosities and extrapolation outside of this distribution. We also check the stability of the models by extrapolation in time.

We find that models that incorporate more physical assumptions result in lower errors on the test set than those without such inductive biases. For example, use of the Boussinesq hypothesis in \textit{transport-II} produces significantly better approximations than \textit{transport-I}, which only models a stress analogue directly. The \textit{resnet} and \textit{anode} models have no assumptions of the PDE form, meaning they must learn all the complex dynamics of the Burgers system with limited training data, resulting in unsuccessful models. Increasing the size of the training set may have allowed for improved learning, however we are specifically interested in data-efficient models given the cost of generating ground truth DNS simulations. At the other extreme, we leverage all the assumptions of the baseline Smagorinsky model and only apply a small modification to learn a spatially varying Smagorinsky constant. Here, we are able to show universal improvement over the baseline Smagorinsky models at all viscosities.

We take our analysis further by demonstrating the resolution invariance of our closure approach. Subgrid closures model flow scales smaller than their cutoff wavenumber, $k=M/2$ in our case. This cutoff should lie in the inertial region of the energy spectra, but where exactly within the region it lies may be variable depending on the coarse grid resolution and system parameters. A universal closure model should be able to adapt to different cutoff wavenumbers, which impacts the range of energy scales it must model. In our second experiment, we examine the adaptive trainable Smagorinsky models in more detail with regards to different coarse grid resolutions. We train the model on four resolutions ranging from cutoffs very close to the dissipative region to cutoffs close to the energy containing region. We show that the model can continue to outperform the constant model at all resolutions, including interpolation between the training resolutions. To the best of our knowledge, this is the first use of a differentiable physics based closure on varying uniform grids.

Finally, we show the impact of two neural network architectural choices -- the size of the hidden layer in the network and the non-locality of the FNO functional form. It is typical in machine learning approaches to create very large networks with 10s of thousands of parameters to create very expressive networks that can act as universal approximators. However, we have included a great deal of inductive bias in our model simply by leveraging the Smagorinsky form, meaning our network only has to learn a small correction to see improvement. We demonstrate that a small network of 520 parameters is enough to learn an effective correction, even showing lower loss than the larger model which can overfit the small training set. Lastly, we see the impact of the non-local FNO by testing a model with linear layers. While indeed the local adaptive model improves over the baseline, the advantage is not as significant as the non-local model, although its computational efficiency is greater.

Closure modeling remains a significant challenge in the fluid dynamics community. Machine learning is enabling a new class of data-driven models, but their performance on arbitrary flows can be difficult to characterize. We use the Burgers system as a test bed for developing subgrid models in a differentiable framework, training on an \textit{a posteriori} loss and testing on a large variety of Burgers systems. We show that a simple correction to existing closure models using data-efficient neural networks can result in better performance.

\newpage
\section{Acknowledgements}
This material is based upon work supported by the National Science Foundation Graduate Research Fellowship under Grant No. DGE 1745016 awarded to VS. The authors from CMU acknowledge the support from the Technologies for Safe and Efficient Transportation University Transportation Center, and Mobility21, A United States Department of Transportation National University Transportation Center. This work was supported in part by Oracle Cloud credits and related resources provided by the Oracle for Research program. RM is supported by the U.S. Department of Energy, Office of Science, Office of Advanced Scientific Computing Research, under Contract~DE-AC02-06CH11357.

\setlength\rotFPtop{-6pt}
\begin{sidewaystable}
    \centering
\caption{Model descriptions}
    \label{tab:models}
\begin{tabularx}{\textwidth}{|r|X|m{72pt}|m{72pt}|m{72pt}|X|}
    \hline
\textbf{Name}  &  \centering\textbf{Equations}  & \centering\textbf{Time-stepper} &  \centering\textbf{Layer type}  &  \centering\textbf{Hidden channels} & \textbf{Notes} \\
    \hline \hline
\textit{none} & 
$\dot{\bar{u}} = \nu\nabla^2\bar u - \bar u\nabla \bar u$ & 
Tsit5 & N/A & N/A & \\ \hline

\textit{smag-const} & 
$\dot{\bar{u}} = \nu\nabla^2\bar u - \bar u\nabla \bar u + \nabla(\nu_t\nabla \bar u)$,\newline
$\nu_t = (C_s\delta x)^2|\nabla \bar u|$ & 
Tsit5 & N/A & N/A & This model contains no neural networks, but $C_s$ is a potentially trainable parameter.\\ \hline 

\textit{smag-dyn} & 
$\dot{\bar{u}} = \nu\nabla^2\bar u - \bar u\nabla \bar u + \nabla(\nu_t\nabla \bar u)$,\newline
$\nu_t = (C_s\delta x)^2|\nabla \bar u|$, \newline
$(C_s\delta x)^2 = \frac{\langle HM \rangle}{\langle M^2 \rangle}$, \newline
$H = \nabla(\widetilde{\bar{u}}^2/2) - \nabla(\widetilde{\bar{u}^2}/2)$, \newline
$M = \kappa^2 \nabla(|\nabla \widetilde{\bar{u}}|\nabla \widetilde{\bar{u}}) - \nabla(\widetilde{|\nabla \bar{u}|\nabla \bar{u}})$ & 
Tsit5 & N/A & N/A & \\ \hline \hline

\textit{resnet} & 
$(\bar{u}_{t+\Delta t}, \eta_{t+\Delta t}) = (\bar{u}_{t}, \eta_{t}) + f_\theta(\bar u_t,\eta_t,\nu;x)\Delta t$, \newline
$\eta_0 = g_{\theta}(\bar u_0; x)$ & 
Euler & FNO & 128 & \\ \hline

\textit{anode} & 
$\dot{\bar{u}}, \dot \eta = f_\theta(\bar u,\eta,\nu;x)$, \newline
$\eta_0 = g_{\theta}(\bar u_0; x)$ & 
Tsit5 & FNO & 128 & \\ \hline

\textit{direct} & 
$\dot{\bar{u}} = \nu\nabla^2\bar u - \bar u\nabla \bar u + \eta$, \newline
$\eta = f_\theta(\nabla \bar u, \nu; x)$ & 
Tsit5 & FNO & 128 & \\ \hline

\textit{transport-I} & 
$\dot{\bar{u}} = \nu\nabla^2\bar u - \bar u\nabla \bar u + \nabla\eta$, \newline
$\dot \eta = \alpha \cdot \nu\nabla^2\eta - \beta \cdot \bar u\nabla\eta$,\newline
$\alpha, \beta = f_\theta(\nabla \bar u, \nabla \eta, \nu; x)$,\newline
$\eta_0 = g_{\theta}(\bar u_0, \nu; x)$ & 
Tsit5 & FNO & 128 & \\ \hline

\textit{transport-II} & 
$\dot{\bar{u}} = \nu\nabla^2\bar u - \bar u\nabla \bar u + \nabla(\eta\nabla \bar u)$, \newline
$\dot \eta = \alpha \cdot \nu\nabla^2\eta - \beta \cdot \bar u\nabla\eta$,\newline
$\alpha, \beta = f_\theta(\nabla \bar u, \nabla \eta, \nu, \delta x; x)$,\newline
$\eta_0 = g_{\theta}(\bar u_0, \nu; x)$ & 
Tsit5 & FNO & 128 & The input $\delta x$ to $f_\theta$ is only used in the resolution study.\\ \hline

\textit{transport-I-p} & 
$\dot{\bar{u}} = \nu\nabla^2\bar u - \bar u\nabla \bar u + \nabla\eta$, \newline
$\dot \eta = \alpha \cdot \nu\nabla^2\eta - \beta \cdot \bar u\nabla\eta + \gamma$,\newline
$\alpha, \beta, \gamma = f_\theta(\nabla \bar u, \nabla \eta, \nu; x)$,\newline
$\eta_0 = g_{\theta}(\bar u_0, \nu; x)$ & 
Tsit5 & FNO & 128 & \\ \hline

\textit{transport-II-p} & 
$\dot{\bar{u}} = \nu\nabla^2\bar u - \bar u\nabla \bar u + \nabla(\eta\nabla \bar u)$, \newline
$\dot \eta = \alpha \cdot \nu\nabla^2\eta - \beta \cdot \bar u\nabla\eta + \gamma$,\newline
$\alpha, \beta, \gamma = f_\theta(\nabla \bar u, \nabla \eta, \nu; x)$,\newline
$\eta_0 = g_{\theta}(\bar u_0, \nu; x)$ & 
Tsit5 & FNO & 128 & \\ \hline

\textit{smag-ml} & 
$\dot{\bar{u}} = \nu\nabla^2\bar u - \bar u\nabla \bar u + \nabla(\nu_t\nabla \bar u)$, \newline
$\nu_t = (\eta\delta x)^2|\nabla \bar u|$,\newline
$\eta = f_\theta(\nabla \bar u, \nu, \delta x; x)$ & 
Tsit5 & FNO & 128 & The input $\delta x$ to $f_\theta$ is only used in the resolution study.\\ \hline

\textit{smag-ml-local} & 
$\dot{\bar{u}} = \nu\nabla^2\bar u - \bar u\nabla \bar u + \nabla(\nu_t\nabla \bar u)$, \newline
$\nu_t = (\eta\delta x)^2|\nabla \bar u|$,\newline
$\eta = f_\theta(\nabla \bar u, \nu, \delta x; x)$ & 
Tsit5 & Linear & 128 & \\ \hline

\textit{smag-ml-small} & 
$\dot{\bar{u}} = \nu\nabla^2\bar u - \bar u\nabla \bar u + \nabla(\nu_t\nabla \bar u)$, \newline
$\nu_t = (\eta\delta x)^2|\nabla \bar u|$,\newline
$\eta = f_\theta(\nabla \bar u, \nu, \delta x; x)$ & 
Tsit5 & FNO & 2 & \\ \hline

\end{tabularx}
\end{sidewaystable}

\clearpage
\bibliography{aipsamp}% Produces the bibliography via BibTeX.
%\addbibresource{aipsamp.bib}

\end{document}

%% file: vpack.tex
\usepackage{amsfonts,amssymb,amsbsy,amsmath}
\usepackage[inline]{enumitem}
\usepackage{physics}
\usepackage{siunitx}       % \SI
\usepackage{cancel}        % \cancel
\usepackage{listings}      % source code formatting

\usepackage{soul}          % \hl
\usepackage{float}         % float positioning option \begin{figure}[H]
\usepackage{graphicx}      % \includegraphics  
\usepackage{caption}
\usepackage{subcaption}
\usepackage{xcolor}
\definecolor{periwinkledark}{RGB}{102, 102, 128}

\usepackage{hyperref} % \autoref

\usepackage{amsthm}
\theoremstyle{definition}

\newcommand{\eqn}[1]{ % numbered equation environment
    \begin{equation}
    \begin{aligned}
        #1
    \end{aligned}
    \end{equation}
}

\newcommand{\vect}[1]{\boldsymbol{#1}}                         % vector field
\newcommand{\defeq}{\mathrel{\mathop:}=}                       % define equal to
\newcommand{\I}{\mathcal{I}}                                   % Operator
\newcommand{\R}{\mathbb{R}}                                    % Set   of reals
\renewcommand{\L}{\mathcal{L}}                                 % Function space
\renewcommand{\bar}[1]{\overline{#1}}